\documentclass[acmsmall,natbib=false]{acmart}

\AtBeginDocument{%
  }

\setcopyright{acmcopyright}
\copyrightyear{2018}
\acmYear{2018}
\acmDOI{XXXXXXX.XXXXXXX}

\acmJournal{JACM}
\acmVolume{37}
\acmNumber{4}
\acmArticle{111}
\acmMonth{8}

\RequirePackage[
  datamodel=acmdatamodel,
  style=acmnumeric,
  ]{biblatex}

\usepackage{graphicx}
\usepackage{textcomp}
\usepackage{xcolor}

\usepackage{latexsym}
\usepackage{url}
\usepackage{listings}

\usepackage[noend]{algpseudocode}

\begin{document}

\title{An Empirical Study of Impact of Solidity Compiler Updates on Vulnerabilities in Ethereum Smart Contracts}

\author{Chihiro Kado}
\authornote{This author contributed to an empirical study, including data collection and analysis, and the paper presentation.}
\email{c-kadou@ist.osaka-u.ac.jp}
\orcid{0009-0006-4400-4638}
\author{Naoto Yanai}
\authornote{This author contributed to data analysis and the paper presentation.}
\email{yanai@ist.osaka-u.ac.jp}
 \orcid{0000-0002-0817-6188}
\author{Jason Paul Cruz}
\authornote{This author contributed to the paper presentation.}
\email{cruz@ist.osaka-u.ac.jp}
\orcid{0000-0002-9935-1534}
\author{Kyosuke Yamashita}
\authornote{This author contributed to the paper presentation.}
\email{yamashita@ist.osaka-u.ac.jp}
\authornotemark[1]
\affiliation{%
  \institution{\\Osaka University}
  \streetaddress{1-5 Yamadaoka}
  \city{Suita}
  \state{Osaka}
  \country{Japan}
  \postcode{565-0871}
}

\author{Shingo Okamura}
\authornote{This author contributed to the paper presentation.}
\email{okamura@info.nara-k.ac.jp}
\affiliation{%
  \institution{National Institute of Technology, Nara College}
  \streetaddress{22 Yatacho}
  \city{Yamatokoriyama}
  \state{Nara}
  \country{Japan}
  \postcode{639-1058}
}

\renewcommand{\shortauthors}{Kado and Yanai, et al.}
\renewcommand{\shorttitle}{An Empirical Study of Impact of Solidity Compiler Updates}

\begin{abstract}
Vulnerabilities of Ethereum smart contracts often cause serious financial damage. 
Whereas the Solidity compiler has been updated to prevent vulnerabilities, its effectiveness has not been revealed so far, to the best of our knowledge. 
In this paper, we shed light on the impact of compiler versions of vulnerabilities of Ethereum smart contracts. 
To this end, we collected 503,572 contracts with Solidity source codes in the Ethereum blockchain and then analyzed their vulnerabilities. 
For three vulnerabilities with high severity, i.e., \textit{Locked Money}, \textit{Using tx.origin}, and \textit{Unchecked Call}, we show that their appearance rates are decreased by virtue of major updates of the Solidity compiler. 
We then found the following four key insights. 
First, after the release of version~0.6, the appearance rate for \textit{Locked Money} has decreased. 
Second, regardless of compiler updates, the appearance rate for \textit{Using tx.origin} is significantly low. 
Third, although the appearance rate for \textit{Unchecked Call} has decreased in version~0.8, it still remains high due to various factors, including code clones. 
Fourth, through analysis of code clones, our promising results show that the appearance rate for \textit{Unchecked Call} can be further decreased by removing the code clones. 
\end{abstract}


\definecolor{verylightgray}{rgb}{.97,.97,.97}

\lstdefinelanguage{Solidity}{
	keywords=[1]{anonymous, assembly, assert, balance, break, call, callcode, case, catch, class, constant, continue, constructor, contract, debugger, default, delegatecall, delete, do, else, emit, event, experimental, export, external, false, finally, for, function, gas, if, implements, import, in, indexed, instanceof, interface, internal, is, length, library, log0, log1, log2, log3, log4, memory, modifier, new, payable, pragma, private, protected, public, pure, push, require, return, returns, revert, selfdestruct, send, solidity, storage, struct, suicide, super, switch, then, this, throw, transfer, true, try, typeof, using, value, view, while, with, addmod, ecrecover, keccak256, mulmod, ripemd160, sha256, sha3}, 
	keywordstyle=[1]\color{blue}\bfseries,
	keywords=[2]{address, bool, byte, bytes, bytes1, bytes2, bytes3, bytes4, bytes5, bytes6, bytes7, bytes8, bytes9, bytes10, bytes11, bytes12, bytes13, bytes14, bytes15, bytes16, bytes17, bytes18, bytes19, bytes20, bytes21, bytes22, bytes23, bytes24, bytes25, bytes26, bytes27, bytes28, bytes29, bytes30, bytes31, bytes32, enum, int, int8, int16, int24, int32, int40, int48, int56, int64, int72, int80, int88, int96, int104, int112, int120, int128, int136, int144, int152, int160, int168, int176, int184, int192, int200, int208, int216, int224, int232, int240, int248, int256, mapping, string, uint, uint8, uint16, uint24, uint32, uint40, uint48, uint56, uint64, uint72, uint80, uint88, uint96, uint104, uint112, uint120, uint128, uint136, uint144, uint152, uint160, uint168, uint176, uint184, uint192, uint200, uint208, uint216, uint224, uint232, uint240, uint248, uint256, var, void, ether, finney, szabo, wei, days, hours, minutes, seconds, weeks, years},	
	keywordstyle=[2]\color{teal}\bfseries,
	keywords=[3]{block, blockhash, coinbase, difficulty, gaslimit, number, timestamp, msg, data, gas, sender, sig, value, now, tx, gasprice, origin},	
	keywordstyle=[3]\color{violet}\bfseries,
	identifierstyle=\color{black},
	sensitive=false,
	comment=[l]{//},
	morecomment=[s]{/*}{*/},
	commentstyle=\color{gray}\ttfamily,
	stringstyle=\color{red}\ttfamily,
	morestring=[b]',
	morestring=[b]"
}

\lstset{
	language=Solidity,
	backgroundcolor=\color{verylightgray},
	extendedchars=true,
	basicstyle=\footnotesize\ttfamily,
	showstringspaces=false,
	showspaces=false,
	numbers=left,
	numberstyle=\footnotesize,
	numbersep=9pt,
	tabsize=2,
	breaklines=true,
	showtabs=false,
	captionpos=b
	}



\received{20 February 2007}
\received[revised]{12 March 2009}
\received[accepted]{5 June 2009}

\maketitle

\section{Introduction} 
\label{sec:Introduction}


    Smart contracts that provide a platform for decentralized applications have attracted attention as applications of blockchains in recent years. 
    Loosely speaking, smart contracts enable automated transactions by storing program code on blockchains, along with information about cryptocurrency transactions. 
    Ethereum \cite{theyellowpaper} is the most popular platform of smart contracts. (Hereafter, we refer to Ethereum smart contracts simply as ``smart contracts.'')

    
    Smart contracts often contain security issues due to the properties of blockchains \cite{zou2019smart}. 
    Program code of smart contracts is stored on the blockchain, and hence it can be analyzed by an adversary due to its transparency, allowing anyone to check the information on the blockchain. 
    Furthermore, the semantics of Solidity, a high-level language for developing smart contracts, is neither formally nor completely specified \cite{hwang2020gapbetweentheoryandpractice}. 
    As a result, many security incidents have been reported involving smart contracts so far\footnote{\url{https://rekt.news}}.
    For example, \$150 million was stolen through the parity-wallets vulnerabilities\footnote{\url{https://techcrunch.com/2017/11/07/a-major-vulnerability-has-frozen-} \url{hundred-of-millions-of-dollars-of-ethereum/}}. 
    In an effort to address the security issues, researchers have developed vulnerability analysis tools \cite{luu2016making,tikhomirov2018smartcheck,feist2019slither,chinen2020hunting,ETHBMC,schneidewind2020ethor,Weiss2019annotary,ashizawa2021eth2vec,kalra2018zeus,tsankov2018securify,mueller2018mythril}. 
    Concurrently, the Solidity compiler has been updated to prevent vulnerable code as another important effort \cite{crafa2019solidity0.5,tantikul2020exploring,hwang2020gapbetweentheoryandpractice,alt2022solcmc}.

    Although the vulnerability analysis tools have been evaluated in various ways \cite{chen2020survey,kushwaha2022systematicreview,durieux2020empirical47587,parizi2018empiricalanalysis,Angelo2019surveytools,kushwaha2022ethereum}, it is unknown whether the updates of the Solidity compiler have improved the security of smart contracts, to the best of our knowledge.
    Solidity is a relatively new language that was introduced in 2015, which could potentially contain vulnerabilities \cite{wang2022realbugfixes}, regardless of the vulnerability analysis tools mentioned above. 
    Despite the fact that compiler updates are essential countermeasures against vulnerabilities, the impact of Solidity compiler updates remains unclear.
    Therefore, to guide research and development in designing practical analysis tools for smart contracts, it is essential to investigate the current status of vulnerabilities in smart contracts from the standpoints of compiler versions. 
    
    In this paper, we answer the following question: \textit{How do Solidity compiler updates affect each vulnerability?} 
    %
    %
    To this end, we first conduct an empirical study for compiler versions. 
    Specifically, we examine how many vulnerable contracts are deployed in the real world.
    We note the above question is \textit{non-trivial}. 
    According to the existing empirical study \cite{wan2021securityperspective}, developers may prefer to use an older version of the compiler, which is more stable, than the latest version since it is often unstable \cite{tantikul2020exploring,crafa2019solidity0.5}. 
    It may indicate that there are more vulnerable contracts than the theoretical estimation \cite{alt2022solcmc} for the compiler updates of Solidity. 
    Although there are several empirical studies \cite{durieux2020empirical47587,perez2021smartcontract} that investigate vulnerable contracts, these studies do not discuss the impact of compiler versions on vulnerabilities. 
    Thus, the existing studies above do NOT imply any insight into the impact of compiler versions on vulnerabilities, which is our key question.

   To answer the above question, we collect 503,572 contracts deployed by 2022/8/31 and then analyze these contracts with respect to vulnerabilities with high severity \cite{tikhomirov2018smartcheck}, i.e., \textit{Locked Money}, \textit{Using tx.origin}, and \textit{Unchecked Call}\footnote{We refer to GitHub for the latest version of severity. \url{https://github.com/smartdec/smartcheck/tree/master/rule_descriptions}}. 
   More specifically, the collected contracts are classified by compiler versions, and then vulnerabilities of these contracts are evaluated by existing vulnerability analysis tool \cite{tikhomirov2018smartcheck}. 
   We then show that \textbf{major updates of the Solidity compiler are effective for the decrease in vulnerabilities} as our main findings. 
   We also found the following four insights: 

    
\begin{enumerate}
\item \textit{Locked money} is significantly decreased after version~0.6 or later versions. 

\item \textit{Using tx.origin} is initially limited; hence, its impact is regarded as minimal independently of the compiler update.

    
    

\item \textit{Unchecked Call} is decreased in version~0.8. 
However, it is slightly high even in the latest compiler versions due to false positives and the cloning of vulnerable code. 



\item When we analyze code clones that are the reuses of existing contracts, we found that the appearance rate for \textit{Unchecked Call} can be further decreased by removing code clones. 

\end{enumerate}

This paper is the full version of our previous work \cite{kado2023empirical} which will be presented at BRAIN 2023. 
In the previous work, we conducted empirical analysis and then showed that the appearance rates of vulnerabilities were decreased by major updates. 
In this paper, we focus on the existence of code clones, that are reuse of existing contracts with minor changes, and then evaluate newly appeared vulnerabilities precisely by removing the code clones. 
We also consider the use of older compiler versions. 
They were future works mentioned in the previous work. 
We currently show that the appearance rates will be more reduced than the results in the previous work. 


\paragraph*{Paper Organization}

The rest of this paper is organized as follows. 
Section~\ref{sec:techincal_background} describes Ethereum smart contracts and related works.
Section~\ref{sec:methodology} presents our methodology for the empirical study, including the underlying tool, vulnerabilities, and data collection.
Section~\ref{sec:results} describes the results of the empirical study, and then Section~\ref{sec:discussion} describes the impacts of compiler versions on the vulnerabilities, including code clones and threats of validity in this paper.
The conclusion and future directions are presented in Section~\ref{sec:conclusion}. 

\section{Background} \label{sec:techincal_background} 

In this section, we describe Ethereum smart contracts and their related works as technical backgrounds. 

\subsection{Ethereum Smart Contracts} \label{sec:ethereum}

Smart contracts in Ethereum are known as an \emph{immutable} computer program that is deployed on the blockchain, and are executed \emph{deterministically} in Ethereum Virtual Machines (EVMs) managed by peers in the Ethereum network. 
The immutability means that the code of smart contracts cannot be changed once deployed. 
On the other hand, the deterministic property means that the execution of functions of the smart contracts will produce the same result even when anyone runs them. 

Smart contracts in Ethereum are implemented with source code written in a high-level language such as Solidity \cite{solidityDoc0.8.0} in general. 
Once deployed on the blockchain, a contract is self-enforcing on EVMs. 
Each smart contract is given a contract address as a unique identifier. 
Using this address, the smart contract can receive Ether, the cryptocurrency of Ethereum, when its contract address is designated as the destination of a transaction transmitted over the network and then its functions are executed. 
When a contract is invoked, the transaction and the execution result are recorded on the blockchain. 
The transaction data also represents the function to be executed and its parameters in the contract. 
To give the peers incentives for executing functions of contracts, Ethereum relies on \emph{gas} as payment of Ether to ``fuel computations". 

Since Vulnerabilities of smart contracts may cause expensive financial damage as found in The DAO hack, their countermeasure tools have been developed so far \cite{chen2020survey}. 
Remarkably, compiler updates can prevent vulnerable code from deployment and execution. 
For instance, execution of code with integer overflow vulnerability fails even when a transaction is given, as long as the code is compiled with the Solidity compiler version 0.8.0 (and later versions).

\subsection{Related Work} \label{sec:relatedworks}

Empirical studies for smart contracts and designs of vulnerability analysis tools are described below. 

\subsubsection{Empirical Study} 
\label{sec:relatedworks_empiricalstudy}

Hwang et al. \cite{hwang2020gapbetweentheoryandpractice} found that 98.14\% of contracts did not have security patches for known vulnerabilities. 
They conducted research until the compiler version~0.5. 
Our work is a subsequent work of their work, and we conduct research until version~0.8. 
Perez et al. \cite{perez2021smartcontract} analyzed the financial damage by vulnerable contracts. 
Combining our work with the above works, it is expected that the financial damage caused by vulnerabilities is estimated for each compiler version. 


Durieux et al. \cite{durieux2020empirical47587} and Ren et al. \cite{ren2021empirical} investigated the performance of existing vulnerability analysis tools with smart contracts, respectively. 
There are also surveys on developments and their release status of vulnerability analysis tools \cite{Angelo2019surveytools,kushwaha2022systematicreview,durieux2020empirical47587,ren2021empirical}. 
According to a questionnaire survey \cite{wan2021securityperspective} for smart contract developers by Wan eta al., only about half of the developers use the latest compiler versions and vulnerability analysis tools. 
Indeed, a new compiler version is often unstable \cite{tantikul2020exploring,crafa2019solidity0.5}, and it may need time until the compiler is stable. 
The surveys described above were conducted from the standpoints of the performance analysis of tools or their developers' viewpoints. 
However, they have not indicated how many vulnerabilities are involved in the real world. 
We believe that it is possible to identify specific vulnerabilities that should be addressed in the future by combining the above works with our work.

\subsubsection{Design of Vulnerability Analysis Tools} 
\label{sec:related-work_Dev.ofTools}

According to existing empirical studies for analysis tools \cite{parizi2018empiricalanalysis,durieux2020empirical47587,kushwaha2022systematicreview}, SmartCheck \cite{tikhomirov2018smartcheck} and Mythril \cite{mueller2018mythril} can provide high accuracy. 
SmartCheck is also faster than Mythril, and hence we utilize SmartCheck for our empirical study. 





\section{Methodology of Empirical Study} 
\label{sec:methodology}

In this section, we describe the research methodology of our empirical study. 
We first describe a vulnerability analysis tool to be utilized in the empirical study and then the preliminary study that was conducted in advance. 
We then describe the empirical study setting, including vulnerabilities to be analyzed, data collection, and metrics. 

\subsection{Vulnerability Analysis Tool}


    We utilize SmartCheck \cite{tikhomirov2018smartcheck} for an empirical study of vulnerabilities. 
    SmartCheck is a tool to analyze vulnerabilities in contracts written in Solidity. 
    Roughly speaking, SmartCheck generates an XML parse tree as an intermediate representation of code and then detects vulnerability patterns by using XPath\footnote{\url{https://www.w3.org/TR/xpath20/}} on the intermediate representation. 
    It can provide a high accuracy \cite{durieux2020empirical47587,parizi2018empiricalanalysis} as described in Section~\ref{sec:relatedworks} and the source code is publicly available via GitHub\footnote{\url{https://github.com/smartdec/smartcheck}}.
    
    We follow the definition by Perez et al. \cite{perez2021smartcontract} as vulnerability: 
    a contract is vulnerable if it has been detected by an analysis tool. 
    It may indicate that some contracts are to be vulnerable due to false positives. 
    According to Perez et al., while false positives are not a serious problem, false negatives need to be reduced possibly. 
    We adopt the above definition in this paper.

    \subsection{Preliminary Study}
        We first conducted a preliminary study on contracts deployed by April 2022. 
        The goal of this preliminary study is to choose compiler versions discussed in this study because several versions may be unstable \cite{tantikul2020exploring,crafa2019solidity0.5}. 

\begin{figure}
            \centering
            \includegraphics[width=\textwidth]{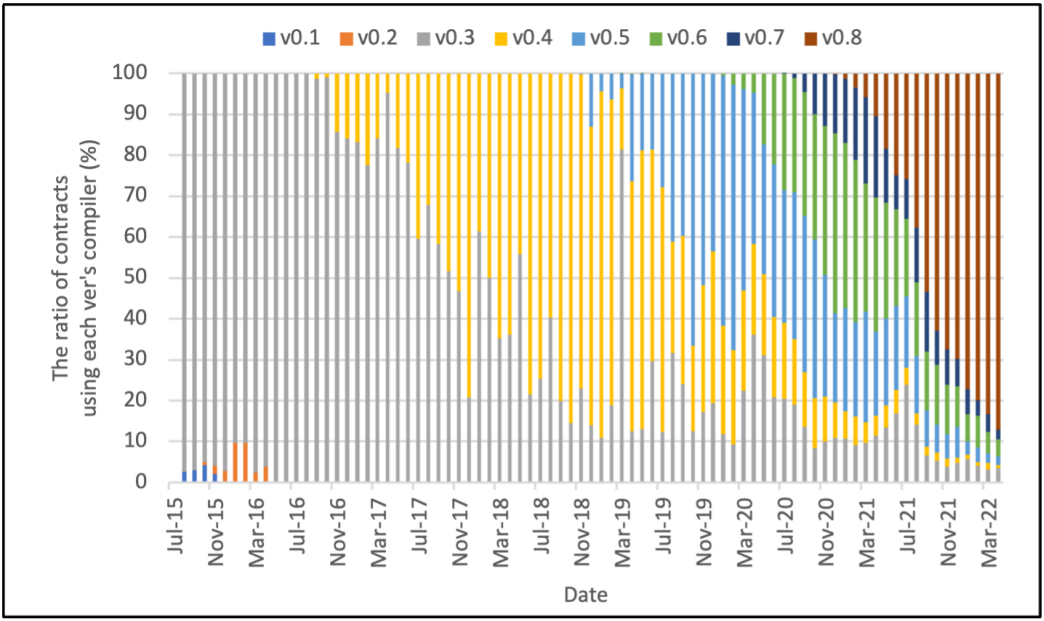}
            \caption{The Ratio of each compiler version. 
            The horizontal axis represents each month and the vertical axis represents the use ratio of contracts complied with each version.}
            \label{fig:contract_total_rel}
        \end{figure}

        We then found only 32 contracts for version~0.1 and 127 contracts for version~0.2. 
        These contracts were deployed by April 2016, and there was no contract compiled with version~0.1 and version~0.2 after that, as shown in Fig.~\ref{fig:contract_total_rel}. 

        It is considered that these versions were seldom used, and this result is consistent with the existing empirical study \cite{pinna2019massiveanalysis}. 
        Meanwhile, there were many empty contracts in version~0.3. 
        These contracts contain only \texttt{0x} as bytecode. 
        They might fail to deploy or be erased from the blockchain. 
        Consequently, we decided that the analysis of version~0.1 to version~0.3 is unstable. 
        Hereafter, version~0.4 and later versions are used for the empirical study. 
        We note that the compiler version~0.4 was released on 2016/9/4. 

        We also focused on vulnerabilities with the highest severity \cite{tikhomirov2018smartcheck} in the preliminary study.
        In general, the severity of vulnerabilities is correlated to the amount of damage caused by them, and hence the analysis based on the highest severity is a more important issue in the real world than other vulnerabilities.
        When we analyzed vulnerabilities with the highest severity, \textit{Constructor Return}, one of those with the highest severity, was not detected. 
        Thus, we analyze three vulnerabilities except for the \textit{Constructor Return} in this study.
        

\subsection{Vulnerabilities to Be Analyzed}
    
    We focus on \textit{Locked Money}, \textit{Using tx.origin}, and \textit{Unchecked Call} as vulnerabilities with the highest severity in our empirical study. 
    We describe each vulnerability in detail below. 
    
    


    \subsubsection{\textit{Locked Money}}
    It was discovered through a parity wallet attack\footnote{\url{https://www.comae.com/posts/the-280m-ethereums-parity-bug./}} whereby a transaction with signatures by multiple users is invalid because one of the signatures is incorrect. 
    It will then cause Ether to be frozen. 
    \textit{Locked Money} occurs when a function is defined to receive Ether with the \texttt{payable} modifier, but there is no function to send Ether, e.g., \texttt{send()} or \texttt{transfer()}.
    The received Ether is then frozen in the contract.
    It is necessary to implement functions to send Ether in order to prevent \textit{Locked Money}. 


    \subsubsection{\textit{Using tx.origin}}
    It occurs due to the global variable, \texttt{tx.origin}, which represents a user's address who sends a transaction. 
    A global variable similar to \texttt{tx.origin} is \texttt{msg.sender}, which represents the caller contract. 
    If \texttt{tx.origin} is used for authentication instead of \texttt{msg.sender}, Ether may be sent to an unauthorized party.


    \subsubsection{\textit{Unchecked Call}}
    
    It is known as \textit{Mishandled Exception} and occurs when a contract calls another contract without considering the exception on the callee \cite{luu2016making}.
    If the callee contract causes the exception when the caller contract sends Ether to the callee, it then rewinds the process of the callee and the callee cannot receive Ether. 
    However, since the caller does not handle the exception, it will lead to lose Ether. 
    In version~0.4.10, the \texttt{transfer()} function was introduced to handle an exception for a caller contract when the exception occurs in a callee contract. 
    It is recommended to use the \texttt{transfer()} function for sending Ether.


\subsection{Setting} 

\subsubsection{Data Collection}

We utilized Etherscan\footnote{\url{https://etherscan.io/}} for data collection of the empirical study. 
Etherscan is a block explorer for Ethereum and provides an API to retrieve information on the blockchain.
We collected all the contracts deployed by August 31, 2022, with Etherscan's API for the empirical study.
We then collected and analyzed 503,572 contracts in total, whose source codes are publicly available and written in Solidity. 

The specific process of the data collection is as follows: 
(1) refer to blocks generated until August 31, 2022; 
(2) search transactions in the blocks whose destinations are empty, i.e., contract-creating transactions, and then obtain their transaction hashes; 
(3) refer to the transaction information using the hash and obtain the address of the created contract;
(4) from the address, obtain the source code and its compiler version information if available. 

\subsubsection{Metrics}

We analyzed the source code by SmartCheck \cite{tikhomirov2018smartcheck} to determine whether it is vulnerable or not. 
We then define two metrics for the empirical study, i.e., \textit{appearance rate} and \textit{monthly appearance rate}. 
They are defined as follows;
%
\begin{eqnarray}
\mbox{Appearance rate} &=&
\frac{\mbox{number of vulnerable contracts}}{\mbox{number of deployed contracts}}.
\end{eqnarray}

The numerator of the above equation represents the number of contracts detected by SmartCheck as vulnerable, and the denominator represents the number of contracts whose Solidity code is publicly available. 
When the above value is computed for each month, we say it the monthly appearance rate. 
Hereafter, we use the above metrics for the empirical study. 

\section{Results of Empirical Study} 
\label{sec:results}

In this section, we demonstrate the results of the empirical study conducted in the setting described in the previous section. 
Specifically, we measure the monthly appearance rates and the appearance rate for each compiler version.




\subsection{Monthly Appearance Rate}

\begin{figure}[b]
    \centering
    \includegraphics[width=\linewidth]{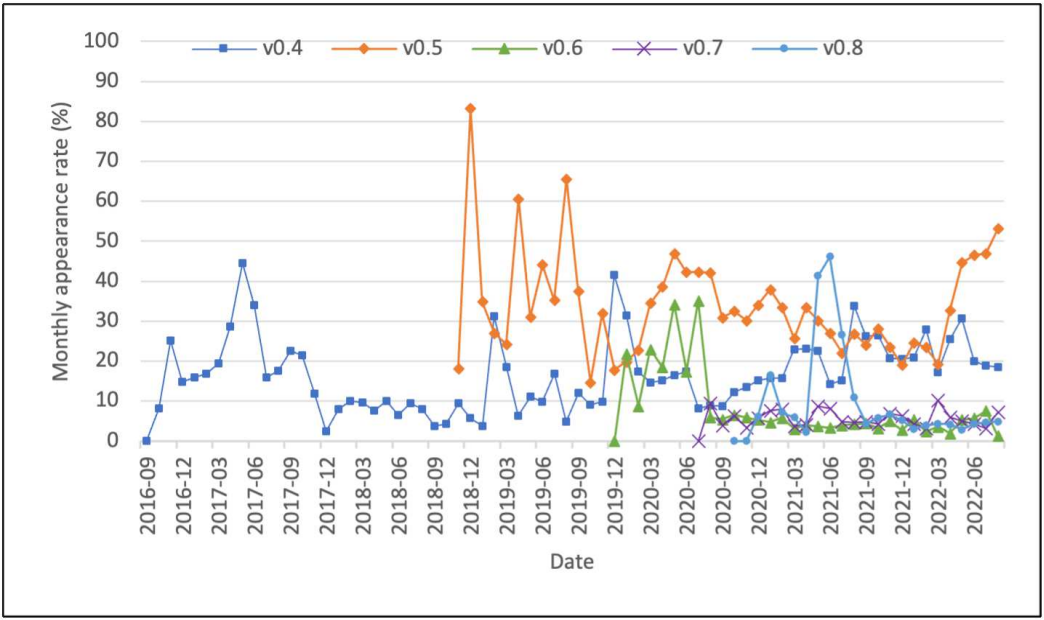}
    \caption{Monthly appearance rates for \textit{Locked Money} in each version.
    }
    \label{fig:locked_money_rel}
\end{figure}

The monthly appearance rates of \textit{Locked Money} for each compiler version are shown in Fig.~\ref{fig:locked_money_rel}. 
Considering the stability of the compiler, we focus on the monthly appearance rates other than version~0.4 and version~0.7 in the period after one year has passed since the latest version, version~0.8, appeared, i.e., from January 2022 to August 2022.
According to the figure, the monthly appearance rate decreased for version~0.6 from version~0.5 during this period.
A similar result as version~0.6 was obtained for version~0.8. 
It means that the monthly appearance rates are almost stable between version~0.6 and version~0.8. 



\begin{figure}[t]
    \centering
    \includegraphics[width=\linewidth]{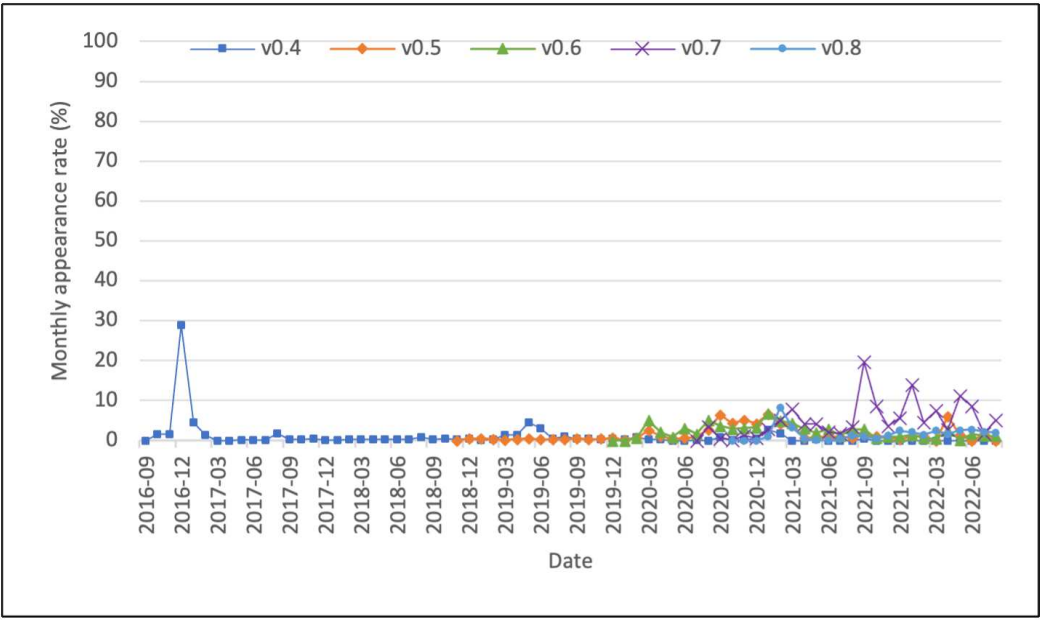}
    \caption{
    Monthly appearance rates for \textit{Using tx.origin} in each version.
    }
    \label{fig:tx_origin_rel}
\end{figure}

Next, the monthly appearance rates of \textit{Using tx.origin} for each compiler version are shown in Fig.~\ref{fig:tx_origin_rel}. 
According to the figure, except for version~0.7, the monthly appearance rates were less than 5\% on average. 
The monthly appearance rates seem to be stable. 



\begin{figure}[ht]
    \centering
    \includegraphics[width=\linewidth]{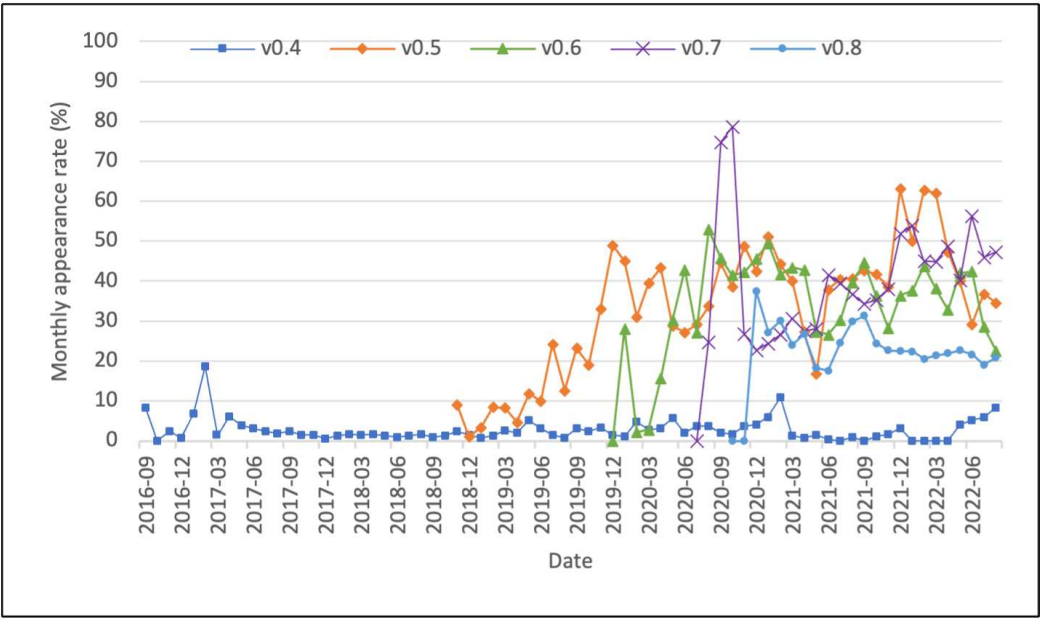}
    \caption{
    Monthly appearance rates for \textit{Unchecked Call} in each version.
    }
    \label{fig:unchecked_call_rel}
\end{figure}

Finally, the monthly appearance rates of \textit{Unchecked Call} for each compiler version are shown in Fig.~\ref{fig:unchecked_call_rel}. 
According to the figure, except for version~0.4 and version~0.7, the monthly appearance rate decreased proportionally to compiler updates in the same period as \textit{Locked Money} except for a small part.
The reason why the monthly appearance rates for version~0.4 and version~0.7 are outliers is described in Section~\ref{sec:v0.4_v0.7}.


\subsection{Appearance Rate}

The appearance rates of the vulnerabilities for each compiler version are shown in Table~\ref{tab:ver_rel}. 
According to the table, the appearance rate has increased for all vulnerabilities after the update from version~0.4 to version~0.5. 
In contrast, looking at the update to version~0.6 and version~0.8, we can observe similar results as those in the previous section, i.e., \textit{Locked Money} and \textit{Using tx.origin} are stable while \textit{Unchecked Call} is decreasing. 
It indicates that similar results will be obtained entirely because of the increase in the number of deployed contracts.

\begin{table}[t]
    \centering
    \caption{Appearance rate for each compiler version. For the columns of vulnerabilities, each value represents the appearance rate, and the value in parentheses represents the number of contracts whose results on SmartCheck are vulnerable, respectively. 
    }
    \begin{tabular}{c|c|c|c|c}
        \hline
        Version & \shortstack{Num. of Deployed \\ Contracts}  & \shortstack{\textit{Locked} \\  \textit{Money}}     & \shortstack{\textit{Using} \\ \textit{tx.origin}}   & \shortstack{\textit{Unchecked}\\ \textit{Call}}\\
        \hline
        version~0.8       & 132912            & \shortstack{6.86\% \\ (9120)}    & \shortstack{1.87\% \\ (2491)} & \shortstack{22.04\% \\ (29297)}   \\ \hline 
        version~0.7       & 22442             & \shortstack{5.73\% \\ (1286)}    & \shortstack{4.90\% \\ (1099)} & \shortstack{38.14\% \\ (8560)}    \\ \hline
        version~0.6       & 54402             & \shortstack{6.66\% \\ (3624)}    & \shortstack{2.97\% \\ (1614)} & \shortstack{49.10\% \\ (21271)}   \\ \hline
        version~0.5       & 83390             & \shortstack{33.60\% \\ (28017)}  & \shortstack{1.84\% \\ (1538)} & \shortstack{32.90\% \\ (27434)}   \\ \hline
        version~0.4       & 210426            & \shortstack{10.66\% \\ (22421)}  & \shortstack{0.61\% \\ (1286)} & \shortstack{1.73\% \\ (3644)}     \\
        \hline
    \end{tabular} \\
    \label{tab:ver_rel}
    \raggedright

\end{table}



\section{Discussion} 
\label{sec:discussion}

In this section, we discuss the impact of compiler updates, including the consideration of version~0.4 and version~0.7 as exceptions.
Next, we discuss \textit{Unchecked Call}, whose monthly appearance rate is high even on the latest version with respect to false positives. 
We also discuss the impact of code clones on the appearance rates for each vulnerability. 
Finally, we discuss threats to the validity of this work, including future work. 



\subsection{Impact of Compiler Updates} \label{sec:impactofupdates}

\subsubsection{Impact of Major Updates} 
\label{sec:major_update}

We discuss the appearance rates for major updates. 
As shown in Section~\ref{sec:results}, since the appearance rate tends to decrease with major updates on the overall scale, it is considered that compiler updates are a contributing factor to the decrease in vulnerabilities. 
We describe the reason based on Table~\ref{tab:ver_rel} below.

First, for \textit{Locked Money}, the appearance rate is significantly decreased by the compiler update from version~0.5 to 0.6: for instance, it is continuously about 6\% after version~0.6. 
Unfortunately, we could not identify the reason for the decrease in the appearance rate. 
Nonetheless, we still consider that the compiler update is effective for the decrease in \textit{Locked Money}. The appearance rate for version~0.5 is still high even after version~0.6 is released, according to Fig.~\ref{fig:locked_money_rel}. 
There were 60,267 contracts for version~0.5 as contracts deployed since December 2019, i.e., version~0.6 was released. 
We also found 18,717 vulnerable contracts for version~0.5 among them with respect to \textit{Locked Money}, i.e., 31.06\% appearance rate. 
Namely, there is a clear gap between version~0.6 (and later versions) and version~0.4/version~0.5. 
We leave it as an open question to deeply understand the decrease in the appearance rate for \textit{Locked Money}. 

Next, for \textit{Using tx.origin}, the appearance rates are lower than 5\% for any version. 
Indeed, avoiding \texttt{tx.origin} for authentication has been described in the Solidity official document \cite{solidityDoc0.8.0}. 
We believe that the description in the official document significantly affects the decrease of \textit{Using tx.origin}. 
We could not confirm the impact of the compiler update for \textit{Using tx.origin} because the number of its vulnerable contracts is quite limited. 
We note that there is no real-world attack for \textit{Using tx.origin} \cite{zhou2022stateofethereum}, and hence the impact of \textit{Using tx.origin} itself is regarded as minimal. 

Finally, for \textit{Unchecked Call}, we focus on version~0.6 or later versions. 
Indeed, although version~0.6 introduced the \texttt{try-catch} syntax, it does not adapt to \textit{Unchecked Call} well \cite{wang2021soliditylanguagefeatures}. 
Furthermore, we focus on only version~0.6 and version~0.8 because version~0.7 is an exception, as described above. 
Then, the appearance rate for \textit{Unchecked Call} is decreased in version~0.8 to less than half of version~0.6. 



Based on the results described above, we conclude as the major updates strikingly affect the decrease in vulnerabilities. 
Although one might think that the appearance rate for \textit{Unchecked Call} is still high even for version~0.8, we discuss this viewpoint in the next subsection in detail. 


\subsubsection{Impact of Minor Updates} 

We discuss the appearance rates for minor updates to investigate the causal relationship between compiler updates and appearance rates in more detail. 
As a result, there were both decreasing and increasing tendencies of appearance rate according to minor updates, and there was no similar change among most of the versions.


\begin{figure}[t]
    \centering
    \includegraphics[width=\linewidth]{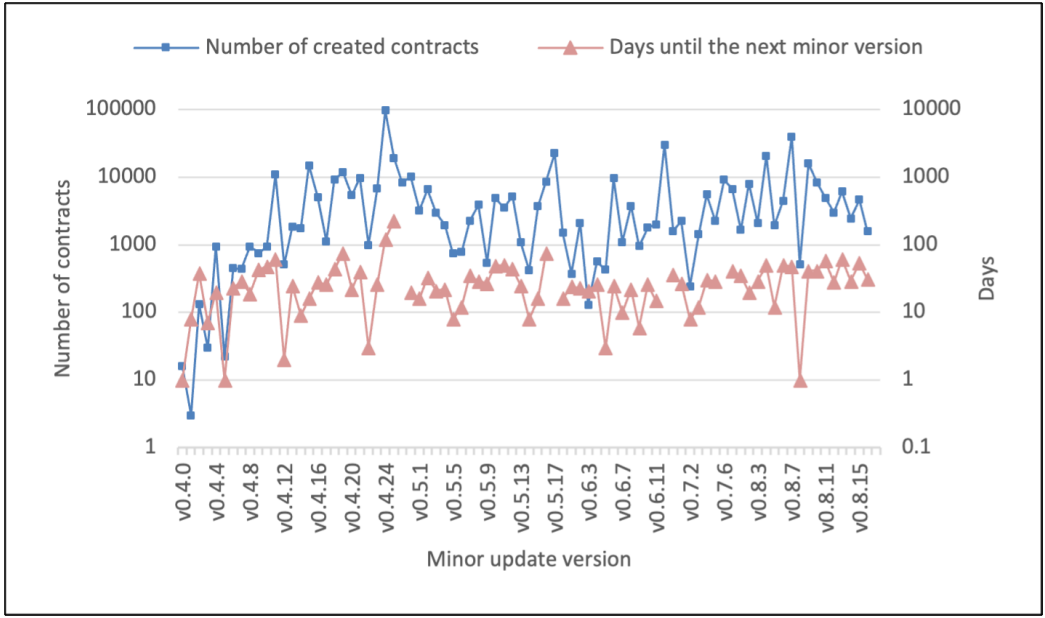}
    \caption{
    The number of deployed contracts and days until the next minor version in each compiler version.
    }
    \label{fig:minor_update}
\end{figure}

It is because the period until the next minor version and the number of deployed contracts vary significantly for each minor version. 
For instance, the period is from one day to more than half a year by the next version, and at least three to at most 97,589 contracts are deployed as shown in Fig.~\ref{fig:minor_update}. 
If the number of deployed contracts is limited, the appearance rate is unstable.
Therefore, it is difficult to confirm the impact of minor updates. 
The compiler update may be executed when a bug is discovered or new language constructions are necessary \cite{pinna2019massiveanalysis}. 
To confirm the detailed factors, we need to investigate the relationship between each vulnerability and the added functions or bugs that have been fixed.
We leave it as an open question to confirm the impact of minor updates. 





\subsubsection{Consideration on version~0.4 and version~0.7}
\label{sec:v0.4_v0.7}

We discuss why version~0.4 and version~0.7 are excluded from the results in Section~\ref{sec:results} below.

First, for version~0.4, there are drastic breaking changes from version~0.4 to version~0.5.
In particular, for sending Ether, we can call any address until version~0.4 and can call only addresses that are \texttt{payable} after version~0.5. 
Furthermore, functions such as \texttt{transfer()} and \texttt{require()} were not implemented until version~0.4.9.  
Since version~0.5 and later versions contain these points, the behavior of version~0.4 and version~0.5 (and later versions) are significantly different. 

Next, version~0.7 has a shorter period until version~0.8 is released compared to the other versions.
The period between version~0.7 and version~0.8 is four and a half months, and the period from version~0.7 to version~0.8.x, the pre-release version of version~0.8, is three months.
It is much shorter than the seven months from version~0.6 to version~0.7, which is the next shortest, and for the other versions, more than a year has passed by the next version.
Even for minor updates, the number of updates is minimal, i.e., only six times for version~0.7, which is half of version~0.6, the next minimal version.
According to the number of deployed contracts shown in Table~\ref{tab:ver_rel}, version~0.7 has been underutilized in less than half of the other versions.
Based on the reasons above, it is considered that version~0.7 stopped providing minor updates before it stabilized.
Therefore, version~0.7 is unstable behavior compared to the other versions.

\subsection{Detailed Analysis of \textit{Unchecked Call}} 
\label{sec:DetailUnchecked}



As shown in Table~\ref{tab:ver_rel}, the appearance rate for \textit{Unchecked Call} decreases to 22.04\% in version~0.8. 
Nevertheless, the decrease in the monthly appearance rate is moderated compared to the other vulnerabilities. 
In this section, We discuss the reason for the high appearance rate for \textit{Unchecked Call} with respect to false positives.
Another reason, Code clone is mentioned in Section~\ref{sec:code_clone}


To identify false positives, we checked the codes detected as vulnerable by SmartCheck. 
We then found the \texttt{delegatecall()} function of the OpenZeppelin library
\footnote{\url{https://github.com/OpenZeppelin/openzeppelin-contracts-upgradeable}}, whose return values are recognized as unchecked by SmartCheck, in several codes. 
We show the code in Listing~\ref{clone_code}.
Since the OpenZeppelin library is known for providing in-depth security \cite{pierro2021analysis} and static analysis tools often have a high false positive rate \cite{ren2021empirical}, we further analyzed the OpenZeppelin library. 
As a result, we confirm that the return values described above are checked by the \texttt{if} statement in line~6 of the code, as shown in Listing~\ref{check_delegatecall}. 
Namely, they did not contain \textit{Unchecked Call} and thus were false positives. 



\begin{figure}[ht]
    \begin{lstlisting}[frame=single,label=clone_code,caption=\texttt{delegatecall()} in OpenZeppelin library,language=Solidity, xleftmargin=16pt]
function _functionDelegateCall(address target, bytes memory data) private returns (bytes memory) {
    require(AddressUpgradeable.isContract(target), "Address: delegate call to non-contract");
    (bool success, bytes memory returndata) = target.delegatecall(data);
    return AddressUpgradeable.verifyCallResult(success, returndata, "Address: low-level delegate call failed");
}
    \end{lstlisting}
\end{figure}
\begin{figure}[ht]
    \begin{lstlisting}[frame=single,label=check_delegatecall,caption=checking \texttt{delegatecall()} in OpenZeppelin library,language=Solidity, xleftmargin=16pt]
function verifyCallResult(
    bool success,
    bytes memory returndata,
    string memory errorMessage
) internal pure returns (bytes memory) {
    if (success) {
        return returndata;
    } else {
        _revert(returndata, errorMessage);
    }
}
    \end{lstlisting}
\end{figure}

\subsection{Impact of Code Clone}
\label{sec:code_clone}

We discuss the impact of code clones on the appearance rates of the vulnerabilities. 
Besides false positives described in the previous subsection, another possible reason for the high appearance rate for \textit{Unchecked Call} is the reuse of source codes because code in smart contracts is often reused as code clones \cite{chen2021understanding}. 
Based on the fact of code clones, we built a hypothesis that the clones of vulnerable codes cause the high appearance rate for \textit{Unchecked Call}. 
We first test the hypothesis as a preliminary analysis. 
After that, based on the hypothesis, we describe how to collect code clones to understand the impact of them on the appearance rates. 
Next, we discuss the impact of code clones on the appearance rates.

\subsubsection{Preliminary Analysis}

We conduct a preliminary analysis to test whether the hypothesis is true. 
Specifically, we first focus on version~0.4 since it has increased from May 2022 as shown in Fig.~\ref{fig:unchecked_call_rel}.

During this period, SmartCheck detected 31 contracts with \textit{Unchecked Call}, while we found that the source codes of 23 contracts are almost the same codes when we manually checked these codes.
It means that 23 contracts among the detected contracts were clones of the same contract. 
Notably, eleven contracts were exactly the same, and the remaining twelve contracts contained one additional function compared to the original contract. 
(Hereafter, we say the original contract if it is unique and the origin of other code clones.)

Based on this result, we believe that the above hypothesis is true. 
Therefore, we explore code clones in more detail as the remaining parts of this subsection.

\subsubsection{Collection of Code Clone}
\label{sec:clone_collection}

We describe how we collect code clones. 
Afterward, we discuss their impact on vulnerabilities and compiler versions used by developers in Section~\ref{sec:clone_consider_uncheckedcall} and Section~\ref{sec:clone_consider_older_version}.
We note that the following results are based on the last six months. 
The original contracts during this period might be clones of those deployed before this period.
It indicates that there might be more code clones than this result.




We define code clones as having all the same code or just adding a few lines with the original contract under the same compiler version. 
It is known as the type-3 clone \cite{Roy07asurvey}. 
We utilize the \texttt{compare} function of the \texttt{difflib.Differ} object in Python3 to identify code clones of Solidity code. 
This function can search for differences between two files and then output added or deleted lines of the files.
We deal with only the contracts, that consist of a single file and are deployed in the last six months, i.e., from March to August 2022.




\subsubsection{Consideration of \textit{Unchecked Call}}
\label{sec:clone_consider_uncheckedcall}

We analyze the contracts collected in Section~\ref{sec:clone_collection} to identify the impact of code clones on the appearance rates of each vulnerability.
In particular, we identify how the appearance rates change by removing the code clones of \textit{Unchecked Call}. 
We also identify the appearance rates of \textit{Locked Money} and \textit{Using tx.origin} by removing code clones for comparison with \textit{Unchecked Call}. 
Hereafter, we call the contracts before removing the code clones as \textit{all contracts} for the sake of convenience. 
%
%
\begin{figure}[ht]
    \centering
    \includegraphics[width=\linewidth]{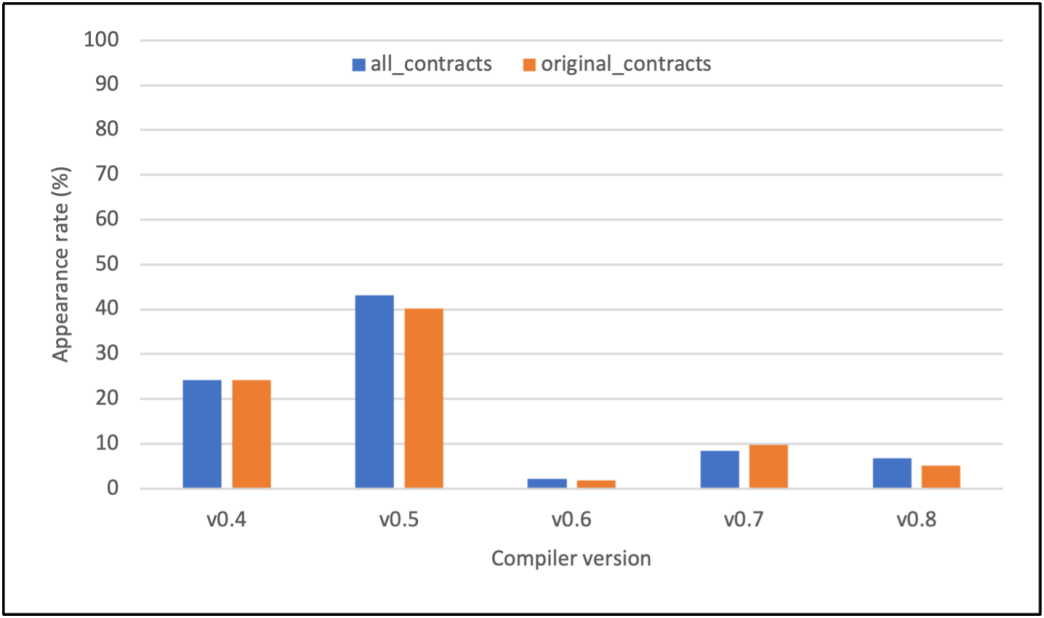}
    \caption{Difference of appearance rates for \textit{Locked Money} between all contracts and original contracts in each version. 
    The all\_contracts contain the all collected contracts, and the original\_contracts contain only the original contracts by removing clones. 
    }
    \label{fig:appearance_rate_LM_Clone}
\end{figure}
\begin{figure}[ht]
    \centering
    \includegraphics[width=\linewidth]{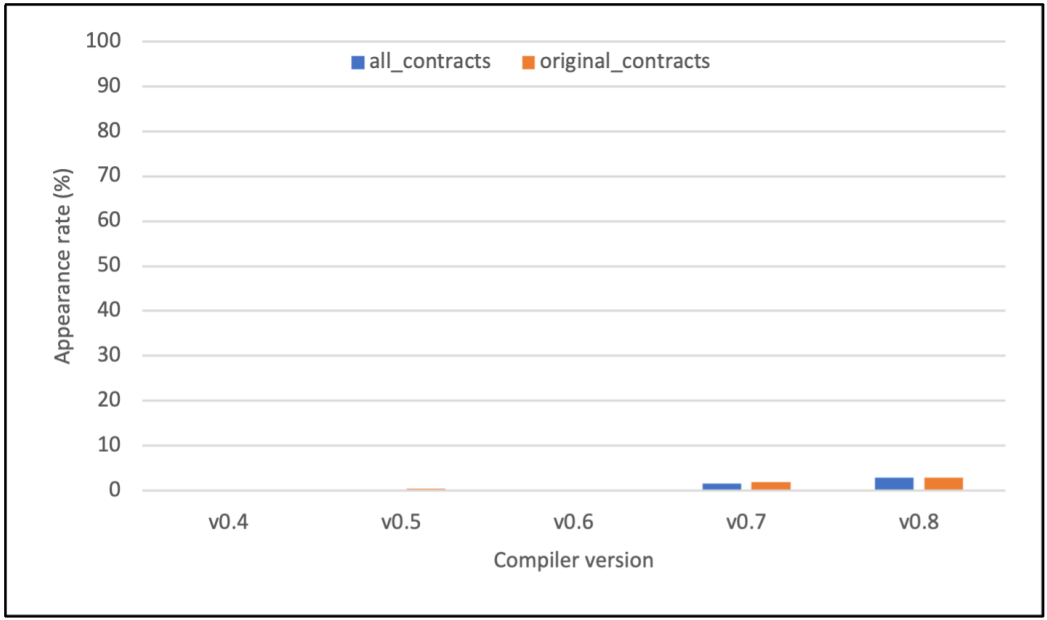}
    \caption{Difference of appearance rates for \textit{Using tx.origin} between all contracts and original contracts in each version. The setting is common with Fig.~\ref{fig:appearance_rate_LM_Clone}. 
    }
    \label{fig:appearance_rate_UT_Clone}
\end{figure}
\begin{figure}[ht]
    \centering
    \includegraphics[width=\linewidth]{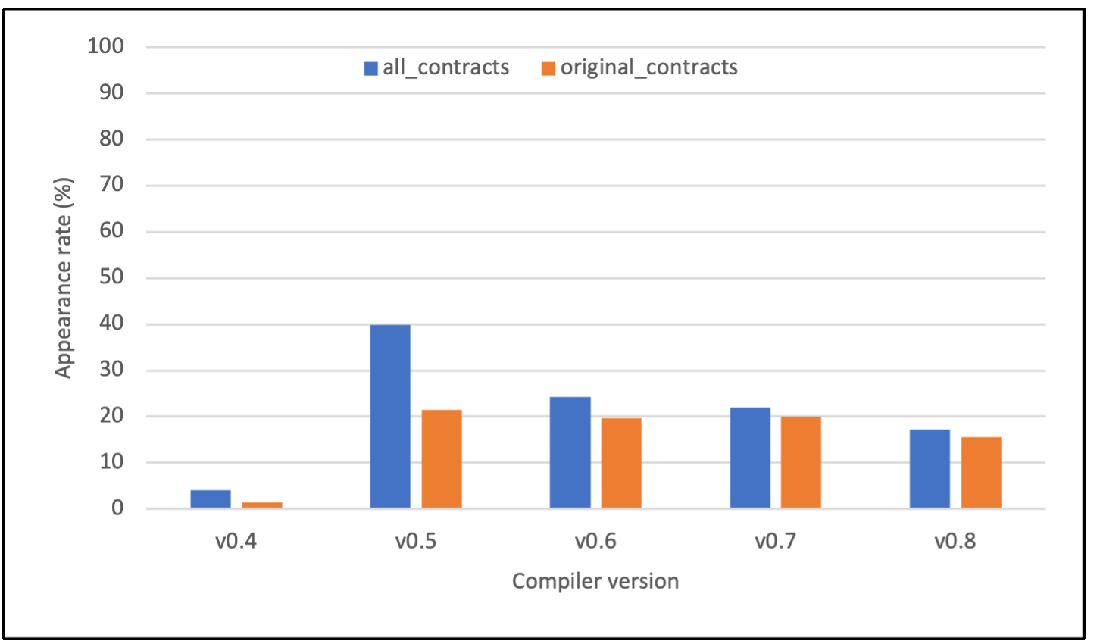}
    \caption{Difference of appearance rates for \textit{Unchecked Call} between all contracts and original contracts in each version. The setting is common with Fig.~\ref{fig:appearance_rate_LM_Clone}. 
    }
    \label{fig:appearance_rate_UC_Clone}
\end{figure}
Figs.~\ref{fig:appearance_rate_LM_Clone}, \ref{fig:appearance_rate_UT_Clone}, and ~\ref{fig:appearance_rate_UC_Clone} show the comparison of the appearance rates for each vulnerability with respect to each compiler version, where all the code clones described above are removed. 

For version~0.7 of \textit{Locked Money} and version~0.5 of \textit{Using tx.origin}, the appearance rates of the original contracts are higher than the all contracts, respectively. 
On the other hand, For \textit{Unchecked Call}, all the appearance rates of the original contracts are lower than the all contracts. 



Here, we focus on the differences between the appearance rates of the all contracts and the original contracts on version~0.8 in comparison between the vulnerabilities. 
We can see that the appearance rates of both \textit{Locked Money} and \textit{Unchecked Call} decrease by removing code clones. 
Based on this fact, as might have been expected, code clones are not a reason why the appearance rate for \textit{Unchecked Call} on version~0.8 is higher than that of the other two vulnerabilities.


Nevertheless, we can still confirm if removing the code clones decreases the appearance rates for \textit{Unchecked Call} in all the versions as shown in Fig. ~\ref{fig:appearance_rate_UC_Clone}. 
In other words, the appearance rates are increased due to the code clones. 
Therefore, for \textit{Unchecked Call}, it is considered that code clones are a reason why the appearance rate is high and it can be decreased by removing the clones. 




\subsubsection{Consideration of the Use of Older Versions}
\label{sec:clone_consider_older_version}


We analyze the use of older compiler versions to provide further evidence for code clones. 
Since compiler updates reduce the vulnerabilities as shown in Section ~\ref{sec:major_update}, using newer compiler versions is strongly recommended. 
However, several contracts have been created using the compiler with older versions, 
and these contracts may contain many code clones: for instance, the appearance rates for the all contracts and the original contracts shown in Fig.~\ref{fig:appearance_rate_UC_Clone} quite differ in version~0.5.  
Consequently, we measure the ratio of the original contracts for each compiler version. 
We show the result in Table~\ref{tab:clone_ratio_for_each_version}.


\begin{table}[ht]
    \centering
    \caption{Details of contracts deployed during the six months}
    \label{tab:clone_ratio_for_each_version}
    \begin{tabular}{c|c|c}
        \hline
        version & Num. of Deployed Contracts & Ratio of Original Contracts \\ \hline
        version~0.4 & 701   & 67.05\% \\   
        version~0.5 & 1771  & 47.15\% \\   
        version~0.6 & 2241  & 81.62\% \\   
        version~0.7 & 716   & 78.91\% \\   
        version~0.8 & 38891 & 91.14\% \\
        \hline
    \end{tabular}
\end{table}


From the table, the ratio of the original contracts over the all contracts tends to increase as a compiler version becomes newer. 
In other words, clones of contracts tend to increase as the compiler version becomes older during the latest six months. 
Besides, the number of the deployed contracts decreases as a compiler version becomes older, except for version~0.7 which is rarely used.
Considering the high ratio of code clones and the number of deployed contracts altogether, we are able to assume that a reason for using older compiler versions is the reuse of contracts created by older versions.


When clones of some contracts are compiled with different compiler versions from the original contracts, the specification change in the Solidity language will lead to errors.
Namely, they cannot be compiled.
Furthermore, developers need to specify the Solidity compiler versions with \texttt{pragma solidity} in their source codes. 
It is written in most of the contracts, and the compiler with the other versions cannot compile them. 
For the reasons described above, we consider that code clones of contracts are compiled with the same version as the original contracts.


However, these clones may have vulnerabilities common with their original contracts as shown in Section~\ref{sec:clone_consider_uncheckedcall} while developers need to create secure contracts.
Consequently, as mentioned at the beginning of this section, when the developers need to reuse the contracts, we recommend the use of a newer compiler version, which fixes errors.

\subsection{Threats to Validity} \label{sec:limitation}


Our empirical study has three threats to validity, i.e., long-term analysis of code clones, false negatives, and developer's skills.
They are issues that need to be addressed in the future. 
The details are described below. 




\subsubsection{Long-Term Analysis of Code Clones}

Our analysis of code clones is limited to the last six months. 
As described in Section~\ref{sec:clone_collection}, the original contracts during this period might be clones deployed in earlier periods. 
Moreover, the number of code clones may differ for periods after each compiler version is released. 
Further studies, which take a long-term analysis of code clones into account, will need to be performed. 

\subsubsection{False Negatives} 

 We only used SmartCheck for the vulnerability analysis of the empirical study.
 It means that the results rely on the analysis accuracy of SmartCheck, and hence there may exist vulnerable codes that are false negatives.
According to the existing study \cite{perez2021smartcontract}, the number of false negatives in analysis can be reduced by introducing more analysis tools and then deciding contracts as vulnerable if any tool decides them as vulnerable. 
Thus, further studies, which introduce more existing analysis tools \cite{mueller2018mythril,luu2016making,feist2019slither,schneidewind2020ethor} other than SmartCheck, will need to be undertaken to reduce false negatives possibly. 

\subsubsection{Developer's Skill}

The developer's coding skills may affect the number of vulnerable contracts, but they were not considered in this paper. 
According to the current questionnaire survey \cite{wan2021securityperspective}, many developers of smart contracts are no longer worried about security, and hence there should exist vulnerable codes due to developer's skills. 
We are in the process of interviewing developers to understand code skills.



\section{Conclusion} 
\label{sec:conclusion}

In this paper, we collected 503,572 contracts deployed by 2022/8/31 with version~0.4 and later version compilers and then conducted the empirical study for vulnerabilities with the highest severity \cite{tikhomirov2018smartcheck}. 
Based on the analysis with SmartCheck, we identified that the appearance rates for \textit{Locked Money}, \textit{Using tx.origin}, and \textit{Unchecked Call} are decreased by virtue of major updates for the Solidity compiler. 
We also found that the appearance rate for \textit{Locked Money} is decreased by about 6\% after version~0.6 while that of \textit{Using tx.origin} is limited initially. 
We also demonstrated that the appearance rate for \textit{Unchecked Call} is decreased by 22.04\% in version~0.8, and it will be further decreased if code clones are removed. 
We plan to conduct extensive evaluations for long-term analysis of code clones and reducing false negatives with more tools \cite{mueller2018mythril,luu2016making,feist2019slither}. 

\begin{acks}
This work was supported in part by MEXT "Innovation Platform for Society 5.0" Program Grant Number JPMXP0518071489 and by JST, CREST Grant Number JPMJCR21M5, Japan.
\end{acks}

\section*{Reproducibility}

Our code is available via GitHub (\url{https://github.com/c-kado/Ethereum_empirical}). 

\printbibliography


\end{document}